\RequirePackage{fix-cm}
\documentclass[12pt]{article}
\usepackage{amsmath,amssymb,amsfonts}
\usepackage{graphicx}

\date{}

\begin{document}

\centerline {\Large{\bf Cold Dark Matter and Holographic Dark Energy \\}}
\centerline{}
\centerline {\Large{\bf  Cosmological Model with Big rip singularity\\ }}

\centerline{}

\centerline{\bf {D.D.Pawar $^{1}$}, S.D.Patil $^{1}$}

\centerline{$^{1}$ School of Mathematical Sciences,}

\centerline{Swami Ramanand Teerth Marathwada University,}

\centerline{Nanded - 431606, (MH), India}

\centerline{E-mail: dypawar@yahoo.com , patil.sona46@mail.com}\
 
\centerline{\bf { V. J. Dagwal * } }
	
\centerline{* Department of Mathematics, Government College of}

\centerline{Engineering, Nagpur- 441 108, India,}

\centerline{E-mail: vdagwal@gmail.com}

\newtheorem{Theorem}{Theorem}[section]

\newtheorem{Definition}[Theorem]{Definition}

\newtheorem{Corollary}[Theorem]{Corollary}

\newtheorem{Lemma}[Theorem]{Lemma}

\newtheorem{Example}[Theorem]{Example}

\begin{abstract}
We have explored cold dark matter and holographic dark energy cosmological model with big rip singularity. To obtain the solution of the field equation, we have supposed that scalar expansion $\theta$ is proportional to shear scalar $\sigma ^2 $  which leads to $p_1=(p_2) ^n $, where $A_1 $, $A_2$ are metric potentials and $ n$ is constant. Big bang and Big rip singularity are investigated. Analyzed the state-finder parameter. The physical and geometrical parameters of the universe are explained.

\end{abstract}

\section{Introduction}
Type - Ia supernovae (SNIeIa) showed that the universe is not expanding at a constant rate; Instead the expansion of the universe is accelerating \cite{1,2,3,4}.
Present astrophysical observations have delivered the astonishing result that dark energy, dark matter and normal matter. More intriguingly, around $70 \%$ of the energy density is in the form of what is called “dark energy”, and is responsible for the acceleration of the distant type Ia supernovae. Hence, today's models of astrophysics and cosmology face two important difficulties, the dark energy problem, and the dark matter problem, respectively\cite{5,6,7,8,9,10}.
Various researchers have studied different types of dark energy such that K-essence \cite{11,12}, quintessence scalar field models \cite{13,14}, Tachyon field \cite{15,16},  Phantom field \cite{17,18}, Quintom field  \cite{19,20}, Chaplygin gas \cite{21,22}, cosmological constant \cite{23, 24}, etc.\\
\par
Einstein's general theory of relativity (GR) was certainly favorable to cosmological models and the evolution of the universe. But Einstein's theory had some drawbacks, e.g. it failed to explain the late time acceleration and accelerated expansion of the universe. Having some limitations in Einstein's theory of general relativity since it does not resolve the problems in cosmology such as dark energy or the lacking matter problem many researchers are attracted towards the alternative theories of gravitation \cite{25,26} \\
\par
 The dark energy universe, known to a large number of astronomers as the holographic dark energy model as it described the dark cosmological region, has attracted astronomical attention \cite{27,28,29}. The holographic dark energy examined by \cite{30,31,32,33}, Has suggested the holographic dark energy universe depends on the holographic principle, the short distance (ultraviolet) cutoff is connected to the long distance (infrared) cutoff it is well-established \cite{34}. Holographic dark energy model have been examined and constrained by distinct astronomical observations \cite{35,36,37}. Saridakis et al \cite{38,39} have examined Tsallis holographic dark energy, which is a generalization of standard holographic dark energy and Barrow holographic dark energy, by applying the ordinary holographic principle at a cosmological framework. Interacting and holographic dark energy model in Bianchi type -I has studied the  Wang et al. \cite{40} respectivily. The study of dark energy is possible during its equation of state (EoS) parameter of holographic dark energy is $\omega_{\Lambda}=\frac{p_\Lambda}{\rho_\Lambda}$ , where ${p_\Lambda} $ is the pressure of holographic dark energy and ${\rho_\Lambda}$ is the energy density of holographic dark energy.\\
 \par
 Recently Pasqua \cite{41} have examined modified holographic dark energy universe in logarithmic f (T) gravity. Dagwal and Pawar \cite{42} have studied Tilted cosmological model consisting two forms of dark energy. Pradhan et al. \cite{43,44} have studied Bianchi type-I anisotropic dark energy model with constant deceleration parameter and Dark energy model in anisotropic Bianchi type-III space-time with variable EoS parameter.\\
 \par 
 The Albert Einstein in 1916,gives the formulation of tensor which is called as general relativity(GR).where as Nordstrom's idea associate gravity with electromagnetism was vanished through the great success of general relativity,it was regenerate ,it imply independently,in 1921 Theoder Kaluza. By using correspondence of field equation of electromagnetism and general relativity.The significant part of Kaluza's work is to determine that the powerful of his five-dimensional theory is similar to the general relativity and electromagnetism.Oskar Klein in 1924 start to thinking resolutely about a fifth dimension which is separate to Kaluza. Same as Nordstrom and Kaluza,Klein was also represent the same as the equation of electromagnetism and gravity.He reflect the Kaluza's theory. Subsequently, different authors studied physics of the universe in the context of higher dimensional space-time. Pawar and Bhuttampalle et al.\cite{45}  have discussed a Kaluza-Klein string cosmological model in the theory of relativity. Also Sahoo and Mishra \cite{46} has been discussed Kaluza-Klein dark energy model in the form of wet dark fluid in f(R,T) gravity.\\
 \par
Motivated by the above discussion, in the present work, we have explored cold dark matter and holographic dark energy cosmological model with big rip singularity. To obtain the solution of the field equation, we have supposed that scalar expansion $\theta$ is proportional to shear scalar $\sigma ^2 $  which leads to $p_1=(p_2) ^n $, where $A_1 $, $A_2$ are metric potentials and $ n$ is constant. Big bang and Big rip singularity are investigated. Analyzed the state-finder parameter. The physical and geometrical parameters of the universe are explained.\\
  This paper is organized as Sect. 2: Metric and Field Equations, Sect. 3: Solution of the field equation, Sect. 4: State-finder Dignostic, Sect. 5:Physical and Geometrical Parameters of the Model in General Relativity, Sect. 6: Results and discussion, Sect. 7: Conclusion.
  
\section{Metric and Field Equations}
We consider the metric in the form
	\begin{equation}\label{1}
		ds^{2}=dt^{2}-P_{1}^{2}(dx^{2}+dy^{2}+dz^{2})-P_{2}^{2}du^{2}
	\end{equation}
	Where $P_{1}, \ P_{2}$ are the function of $t$ only and $u$ is space like\\
The field equation of cold dark matter and holographic dark energy is given by
\begin{equation}\label{2}
R_{ab}-\frac{1}{2}g_{ab}R=-({^m}{T}{_{ab}}+
{^\Lambda}{T}{_{ab}}) .
\end{equation}
The energy momentum tensor for cold dark matter and holographic dark energy are given by\\

\begin{equation}\label{3}
 {{^m}{T}{_{ab}}=\rho_{m} u_{a}u_{b} } \,\,\,\ and \,\,\,\, {^\Lambda}{T}{_{ab}}=(\rho_{\scriptscriptstyle\Lambda}+p_{\scriptscriptstyle\Lambda})u_{a}u_{b}-g_{ab}p_{\scriptscriptstyle\Lambda}
\end{equation}
 where $\rho_m$ , $\rho_{\scriptscriptstyle\Lambda}$ are the energy density of cold dark matter, holographic dark energy and  $p_{\scriptscriptstyle\Lambda}$ is the pressure of holographic dark energy. 
 \par
The field equations are given by

\begin{align}
\frac{2\ddot{P_{1}}}{P_{1}}+\frac{\dot{P_{1}^{2}}}{P_{1}^{2}}+\frac{2\dot{P_{1}}\dot{P_{2}}}{P_{1}P_{2}}+\frac{\ddot{P_{2}}}{P_{2}} & = -p_{\scriptscriptstyle\Lambda}   \label{4}\\ 
\frac{3\ddot{P_{1}}}{P_{1}}+\frac{3\dot{P_{1}^{2}}}{P_{1}^{2}} & = -p_{\scriptscriptstyle\Lambda}\label{5} \\
\frac{3\dot{P_{1}^{2}}}{P_{1}^{2}}+\frac{3\dot{P_{1}}\dot{P_{2}}}{P_{1}P_{2}}& = \rho_{m}+\rho_{\scriptscriptstyle\Lambda}\label{6}
\end{align}	
 Where overhead $(\cdot)$ represent derivative with respect to cosmic time $t$.
  \par
In order to initiate that interaction between energy momentum tensor for cold dark matter and holographic dark energy. Here we suppose that these two components don't conserve separately but interconnect with each other.  The equilibrium equations are given as

\begin{align}\label{7}
		\dot{\rho_{m}} + \frac{\dot{V}}{V} \rho_{m} & = Q
	\end{align}
	\begin{align}\label{8}
		\dot{\rho_{\scriptscriptstyle\Lambda}} + \frac{\dot{V}}{V}\left(1+\omega_{\scriptscriptstyle\Lambda}\right)\rho_{\scriptscriptstyle\Lambda} & = -Q
	\end{align}

Where, $\displaystyle \omega_{\scriptscriptstyle\Lambda}=\frac{p_{\scriptscriptstyle\Lambda}}{\rho_{\scriptscriptstyle\Lambda}}$ is equation of state parameter of holographic dark energy and $Q$ is the interacting term and $Q>0$ measure the stability of the interaction. $Q$ implies that cold dark matter and holographic dark energy conserve separately. The interaction matter dark energy of featuring model is introduced by,  Billyard and Coley \cite{47}, Wetterich \cite{48,49} and Horvat \cite{50}is used the holographic dark energy. In prospect of the continuity equation, the interplay among DE and DM should be the role of the energy density proliferated by a number with units of inverted of time, which may be selected as the Hubble component $H$.
	For any consolidation of dark energy  and dark matter which is freedom to take the form of energy density such that the interplay among DE and DM should be considered intentional in form like as in \cite{51,52,53}. \\
We have
	\begin{equation}\label{9}
		Q= 4b^{2}H\rho_{m}=b^{2}\frac{\dot{V}}{V}\rho_{m} 
	\end{equation}
	Where, $b^{2}$ is coupling constant.
	\par
	Li et al.\cite{54} have occupied the same connection for interacting dark matter and phantom dark energy in procedure to escape the coincidence problem with the help of equation (\ref{7}) and (\ref{9}). We get the energy density  (DM) like
\begin{align}\label{10}
		\rho_{m} & = \rho_{0}V^{(b^{2}-1)} ,
	\end{align}
	where $\rho_{0}>0$ is the integrating constant.\\
	 With the help of equation (\ref{9}) and (\ref{10}) we obtain the interacting term $Q$ as
	\begin{equation}\label{11}
		Q=4\rho_{0}b^{2}HV^{(b^{2}-1)}
	\end{equation}
	\section{Solution of the field equation}
Here the system of equations (\ref{4})-(\ref{6}), reduces to three independent equations with five unknowns $P_1$, $P_2$, $\rho_m$ , $\rho_{\scriptscriptstyle\Lambda}$ and $p_{\scriptscriptstyle\Lambda}$. So we required some physical significant conditions.
	\begin{equation}\label{12}
P_{1}   = P_{2}^{n}
	\end{equation}
where $n>1$ is constant.\\
From equations (\ref{4}) and (\ref{5}) , we get

\begin{equation}\label{13}
		\frac{\ddot{P_{1}}}{P_{1}}+\frac{2\dot{P_{1}^{2}}}{P_{1}^{2}}-\frac{2\dot{P_{1}}\dot{P_{2}}}{P_{1}P_{2}}-\frac{\ddot{P_{2}}}{P_{2}}=0
	\end{equation}
From equations (\ref{12}) and (\ref{13}), we get
	\begin{equation}\label{14}
P_{1}  = (3n+1)^{\frac{n}{3n+1}}(k_{1}t+k_{2})^{\frac{n}{3n+1}}
 	\end{equation}
	\begin{equation}\label{15}
		P_{2} = (3n+1)^{\frac{1}{3n+1}}(k_{1}t+k_{2})^{\frac{1}{3n+1}}
	\end{equation}
	Where, $k_{1}>0, \ k_{2}$ is the integrating constant.\\
	 The spatial Volume as
	\begin{equation}\label{16}
		V= P_{1}^{3}P_{2}
	\end{equation}
	\begin{equation}\label{17}
		V = (3n+1)(k_{1}t+k_{2})
	\end{equation}
Using equation (\ref{17}) in equations (\ref{10}) and (\ref{11}) we get
 	\begin{equation}\label{18}
		\rho_{m} =\rho_{0}(3n+1)^{(b^{2}-1)}(k_{1}t+k_{2})^{(b^{2}-1)}
	\end{equation}
	 
	\begin{equation}\label{19}
		Q = b^{2}k_{1}\rho_{0}(3n+1)^{(b^{2}-1)}(k_{1}t+k_{2})^{(b^{2}-1)}
	\end{equation}
 The energy density of holographic dark energy is given by\\
\begin{equation}\label{20}
\rho_{\scriptscriptstyle\Lambda}  =\frac{3nk_{1}^{2}(n+1)}{(3n+1)^{2}(k_{1}t+k_{2})^{2}}-\rho_{0}(3n+1)^{(b^{2}-1)} (k_{1}t+k_{2})^{(b^{2}-1)}
\end{equation}
 The pressure of holographic dark energy is given by 
	\begin{equation}\label{21}
p_{\scriptscriptstyle\Lambda}  = \frac{3nk_{1}^{2}(n+1)}{(3n+1)^{2}(k_{1}t+k_{2})^{2}}
	\end{equation}
The equation of state as 
\begin{equation}\label{22}
		\omega_{\scriptscriptstyle\Lambda}=\frac{p_{\scriptscriptstyle\Lambda}}{\rho_{\scriptscriptstyle\Lambda}} = \frac{\frac{3nk_{1}^{2}(n+1)}{(3n+1)^{2}(k_{1}t+k_{2})^{2}}}{\frac{3nk_{1}^{2}(n+1)}{(3n+1)^{2}(k_{1}t+k_{2})^{2}}-\rho_{0}(3n+1)^{(b^{2}-1)} (k_{1}t+k_{2})^{(b^{2}-1)}}
\end{equation}

\begin{figure}[htbp]
	\includegraphics[width=0.8 \textwidth]{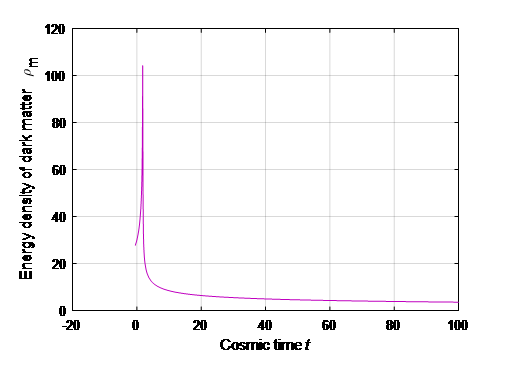}
	\centering 
	\caption { Evolution of energy density of dark matter $\rho_{m}$ with respect to  cosmic time $t$ for $\rho_{0}= 1.2, n = 0.2, b =-2.9, k_1 = -1.14 , k_2 = 2.11  $. }
		\label{fig 1}
		
\end{figure}

\begin{figure}[htbp]
	\includegraphics[width=0.8\textwidth]{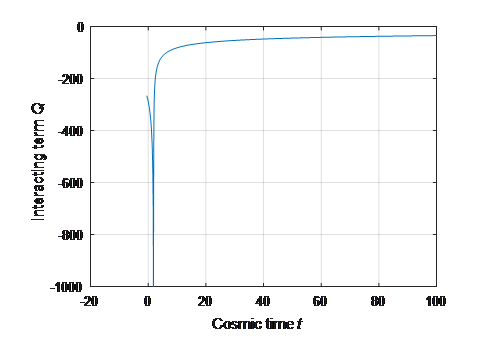}
	\centering 
	\caption { Evolution of interacting term $Q $ with respect to  cosmic time $t$ for $\rho_{0}= 1.2, n = 0.2, b =-2.9, k_1 = -1.14 , k_2 = 2.11  $. }
	\label{fig 2}
	
\end{figure}

\begin{figure}[htbp]
	\includegraphics[width=0.9\textwidth]{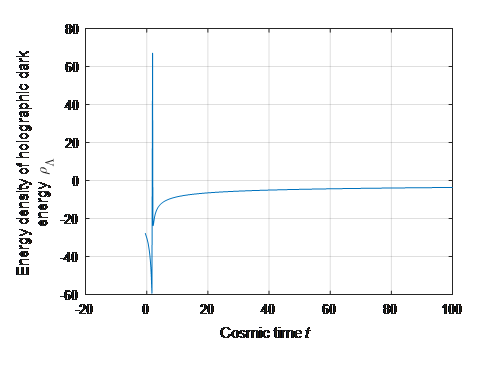}
	\centering 
	\caption { Evolution of energy density of holographic dark energy $\rho_{\scriptscriptstyle\Lambda}$  with respect to  cosmic time $t$ for $\rho_{0}= 1.2, n = 0.2, b =-2.9, k_1 = -1.14 , k_2 = 2.11  $. }
	\label{fig3}
	
\end{figure}

\begin{figure}[htbp]
	\includegraphics[width=0.8\textwidth]{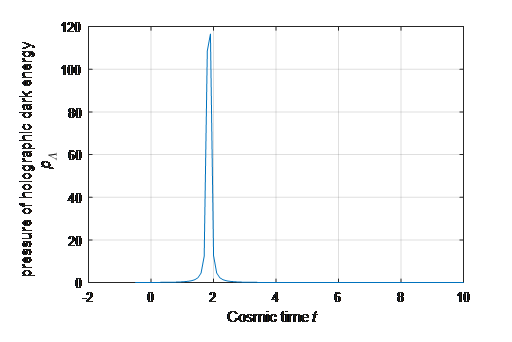}
	\centering 
	\caption { Evolution of pressure of holographic dark energy $p_{\scriptscriptstyle\Lambda}$  with respect to  cosmic time $t$ for $ n = 0.2, k_1 = -1.14 , k_2 = 2.11  $. }
	\label{fig4}
	
\end{figure}

\begin{figure}[htbp]
	\includegraphics[width=0.8\textwidth]{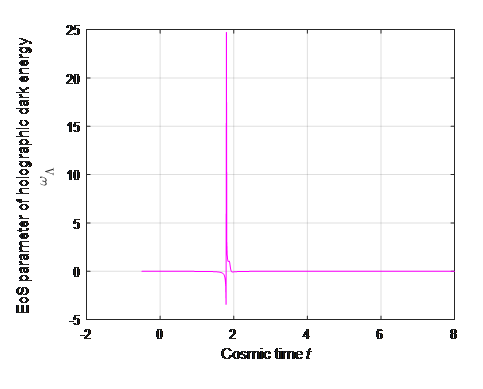}
	\centering 
	\caption { Evolution of  EoS parameter of holographic dark energy $\omega_{\scriptscriptstyle\Lambda} $ with respect to  cosmic time $t$ for  $\rho_{0}= 1.2, n = 0.2, b =-2.9, k_1 = -1.14 , k_2 = 2.11  $. }
	\label{fig 5}
	
\end{figure}

\section{State-finder Dignostic}
There are various DE models have been formed for explaining the cosmic acceleration, the abundance of a different abstract model of dark energy. A sensitive and powerful diagnostic for dark energy model is must for this determination the author Sahni et al. \cite{55} have introduced new mathematical parameter pair $\{r,s \}$, so is known as "state-finder diagnostic".\\
$~~~$	The state-finder parameter $\{\text{r },\text{ s} \}$ is defined(in terms of $H$ and $q$) as follows
	\begin{equation*}
		r=\frac{\dddot{a}}{aH^{3}} \qquad \ \text{and } \quad S= \frac{r-1}{3\left(q-\frac{1}{2} \right)}
	\end{equation*}
Where $a$ is average scale factor and $q$ is the deceleration parameter.
\par
It's important property is that $\{r,s \} = \{ 1,0\}$ is fixed point in the diagram, for the flat $\Lambda$CDM cosmological model. In this paper, the state-finder parameter of DE models have successfully differentiated inclusive the Chaplygin gas, the cosmological constant, quintessence and interacting DE universe \cite{52,54,56}.
\begin{equation}\label{23}
	  r = 21  \  \text{and } \ S= \frac{r}{8} 
\end{equation}
\section{Physical and Geometrical Parameters of the Model in General Relativity}
 In this section we have studied the geometrical and physical parameter of the model.\\  
  The volume scale factor($V$) are given as 
  \begin{align}\label{24}
    V  = \sqrt{g} = P_{1}^{3}P_{2}\qquad \quad \quad (\because g=P_{1}^{6}P_{2}^{2})
  \end{align}
  The average scale factor is defined as 
  \begin{align}\label{25}
    a = V^{\frac{1}{4}}= (P_{1}^{3}P_{2})^{\frac{1}{4}}
  \end{align}
The mean Hubble parameter($H$) is defined  as
\begin{align}
H   = \frac{1}{4}\sum_{i=1}^{4}H_{i}\nonumber \nonumber
\end{align}
It is found to be equal to
   \begin{align}\label{26}
H  = \frac{1}{4}\left[\frac{k_{1}}{(k_{1}t+k_{2})} \right]
\end{align}
The mean anisotropy parameter($\Delta$) is defined  as 
\begin{equation*}
\Delta =\frac{1}{4}\sum_{i=1}^{4}\left(\frac{H_{i}-H}{H} \right)^{2}
\end{equation*} and found as
   \begin{align}\label{27}
\Delta & =\frac{3(n-1)^{2}}{(3n+1)^{2}}
\end{align}
where, $\displaystyle H_{1}=H_{2}=H_{3}=\frac{\dot{P_{1}}}{P_{1}}$ and $\displaystyle H_{4}=\frac{\dot{P_{2}}}{P_{2}}$\\
The coincidence parameter $\displaystyle \overline{r}$  is obtained as 
\begin{equation}\label{28}
\overline{r}=\frac{\rho_{m}}{\rho_{\scriptscriptstyle\Lambda}}= \frac{\rho_{0}(3n+1)^{(b^{2}-1)}(k_{1}t+k_{2})^{(b^{2}-1)}}{\frac{3nk_{1}^{2}(n+1)}{(3n+1)^{2}(k_{1}t+k_{2})^{2}}-\rho_{0}(3n+1)^{(b^{2}-1)}(k_{1}t+k_{2})^{(b^{2}-1)}}
\end{equation}       
The Deceleration parameter $(q)$ is defined as 
\begin{equation}\label{29}
 q=\frac{d}{dt}\left(\frac{1}{H}\right)-1
\end{equation}       and evaluated as
$q=3$

\begin{figure}[htbp]
	\includegraphics[width=0.8\textwidth]{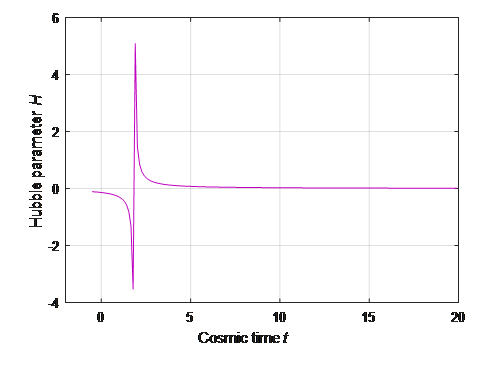}
	\centering 
	\caption { Evolution of Hubble parameter($H$)  with respect to  cosmic time $t$ for $  k_1 = -1.14 , k_2 = 2.11  $. }
	\label{fig 6}
	\includegraphics[width=0.8\textwidth]{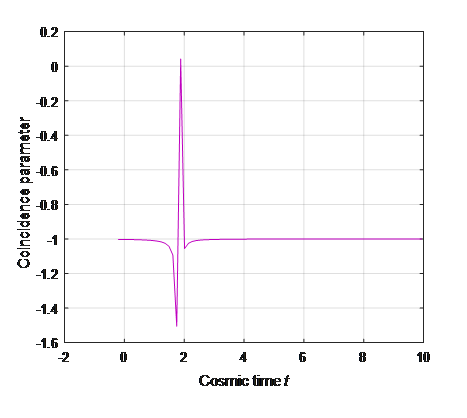}
	\centering 
	\caption { Evolution of coincidence parameter $\displaystyle \overline{r}$  with respect to  cosmic time $t$ for  $\rho_{0}= 1.2, n = 0.2, b =-2.9, k_1 = -1.14 , k_2 = 2.11  $. }
	\label{fig 7}
\end{figure}

\section{Results and discussion}
 Figure 1 represents the  evolution of energy density of dark matter $\rho_{m}$ with respect to  cosmic time $t$ by  taking the values of $\rho_{0}= 1.2, n = 0.2, b =-2.9, k_1 = -1.14 , k_2 = 2.11  $. Equation (18), shows the energy density of dark matter $\rho_{m}$. Its depends on $b^2 $, when $b^2 \rightarrow  1$, the energy density of dark matter $\rho_{m}$ is constant and it has infinite at $ t \rightarrow -k_2/k_1 $ and $b^2  < -1 $. The energy density of dark matter $\rho_{m} $ disappears for $ t \rightarrow \infty $ and $b^2  < -1 $.\\
The profile of interacting term $Q $ with respect to  $t$ is presented in Figure 2 by  taking the range of $\rho_{0}= 1.2, n = 0.2, b =-2.9, k_1 = -1.14 , k_2 = 2.11  $. The interacting term $Q $ has negative value shown in Figure 2 . Equation (19), represents the interacting term $Q $ and same as above results. The interacting term $Q $ has big bang at $ t \rightarrow \infty $ and $b^2  < -1 $ and big rip at $ t \rightarrow -k_2/k_1 $ and $b^2  < -1 $.\\
The progression of the energy density of holographic dark energy $\rho_{\scriptscriptstyle\Lambda}$  with respect to  cosmic time $t$ is  represented  in Figure 3 by  selecting the range of $\rho_{0}= 1.2, n = 0.2, b =-2.9, k_1 = -1.14 , k_2 = 2.11  $. The  $\rho_{\scriptscriptstyle\Lambda}$ is increasing with constant negative value against cosmic time $t$ . In equation (20) shows the energy density of holographic dark energy $\rho_{\scriptscriptstyle\Lambda}$ is diverges when $ t \rightarrow -k_2/k_1 $, $b^2  < -1 $  and disappear at $ t \rightarrow \infty $ , $b^2  < -1 $. The  $\rho_{\scriptscriptstyle\Lambda}$ has singularity between big bang and big rip. \\ 
The evolution of pressure of holographic dark energy $p_{\scriptscriptstyle\Lambda}$  with respect to  cosmic time $t$ is shows in Figure 4 by  taking the values of  $ n = 0.2, k_1 = -1.14 , k_2 = 2.11 $.  Equation (21) represents  the $p_{\scriptscriptstyle\Lambda} \rightarrow 0$ when $ t \rightarrow \infty $ . The  pressure of holographic dark energy stop for  $ n =-1 $.  it is diverges at $ t = -k_2/k_1 $  and disappears for big value of cosmic time $t$. The pressure of holographic dark energy  has intermediate phase between  big bang singularity for big value of cosmic time $t$  and  big rip at $ t = -k_2/k_1 $ .
The variation of the EoS parameter of holographic dark energy $\omega_{\scriptscriptstyle\Lambda} $ against cosmic time $t$  is presented in Figure 5 by  taking the values off $\rho_{0}= 1.2, n = 0.2, b =-2.9, k_1 = -1.14 , k_2 = 2.11  $. 
Figure 5 shows that in initial stage of evolution of the model, the EoS parameter of holographic dark energy $\omega_{\scriptscriptstyle\Lambda} $ was positive therefore  the model was matter dominated. This model  is evolving with negative value at the current time.\\
The progression of the Hubble parameter $H$  with respect to  cosmic time $t$ is shown in Figure 6 by  selecting the values of $  k_1 = -1.14 , k_2 = 2.11  $. The Hubble parameter($H$) is disappears at big value of cosmic time $t$ and it is diverges for $ t \rightarrow -k_2/k_1 $ . 
 The evolution of coincidence parameter $\displaystyle \overline{r}$  against  $t$ is  represented  in Figure 7  by  selecting the range of  $\rho_{0}= 1.2, n = 0.2, b =-2.9, k_1 = -1.14 , k_2 = 2.11  $. The mean anisotropy parameter $\Delta$ and the Deceleration parameter $(q)$ is constant.

\section{ Conclusion}
In this work we have investigated the cold dark matter and holographic dark energy cosmological model with big rip singularity. To obtain the solution of the field equation, we have supposed that scalar expansion $\theta$ is proportional to shear scalar $\sigma ^2 $  which leads to $p_1=(p_2) ^n $, where $A_1 $, $A_2$ are metric potentials and $ n$ is constant. Here we discussed the Eos $\omega_{\Lambda}<0$ is inevitably maintained by the decay of the DE model into pressure-less dark matter $(b^2>0).$ Also, it is investigated that for a suitable option of networking between dark matter and holographic dark energy with $ b^2 = 1$. The energy density of dark matter $\rho_{m} $ , the interacting term $Q $ and the energy density of holographic dark energy $\rho_{\scriptscriptstyle\Lambda}$  have intermediate phase between big bang at $ t \rightarrow \infty $ and $b^2  < -1 $ and big rip at $ t \rightarrow -k_2/k_1 $ and $b^2  < -1 $. The pressure of holographic dark energy and the Hubble parameter $H$  have big bang singularity  when $ t \rightarrow \infty $ and it has big rip singularity when $ t \rightarrow -k_2/k_1 $.

\begin{figure}[htbp]
\includegraphics[width=0.6\textwidth]{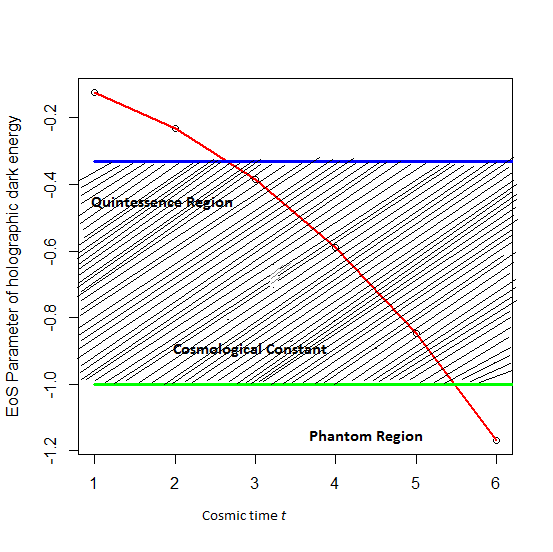}
\centering 
\caption{ Evolution of  EoS parameter of holographic dark energy $\omega_{\scriptscriptstyle\Lambda} $ with respect to  cosmic time $t$}\label{fig1}
\end{figure}

Figure 5 and \ref{fig1} shows that the evolution of EoS parameter $\omega_{\Lambda}$ with cosmic time $t$ by  taking the values of  $\rho_{0}= 1.2, n = 0.2, b =-2.9, k_1 = -1.14 , k_2 = 2.11  $. In Figure 5 and Figure 8 , we have observed that at the initial time $\omega_{\Lambda}$ enters into quintessence region $(-1<\omega <-1/3)$ and at the late time it attains the cosomological consatant $(\omega_{\Lambda}=-1)$ and $\omega_{\Lambda}$ infinite in the phantom model $(\omega_{\Lambda}<-1)$. 
The  parameters $\Delta$ and  $ (q) $ are constant.
In Figure 7 , the coincidence parameter $\displaystyle \overline{r}$ is  decreases against  $t$ and tends to a constant negative range in the big-time limit.

\begin{figure}[htbp]
\includegraphics[width=0.7\textwidth]{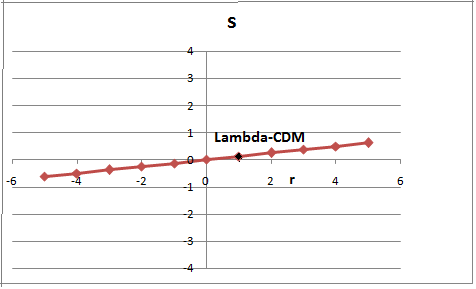}
\centering 
\caption{Evolution of  state-finder parameter s verses r}\label{fig2}
\end{figure}

The state-finder parameter is correlated to the universe in order to distinguish between the dark energy universe with other present dark energy universes. Figure \ref{fig2} shows that in the $s-r$ plane for correspondent models is distinct from those of further dark energy models. The state-finder parameter line passes the point $(s,r)=(0,1)$ which correspondence to the $\Lambda$CDM model. The $s-r$ plane provide the region of phantom $(s>0)$, quintessence region $(r<1)$ DE models.This result agrees with Jawad and  Chattopadhyay \cite{57}.

\newpage
\section{Acknowledgements}
The authors are grateful to the anonymous refree for his/her voluable comments and suggestions which helped to improve the quality of research paper.


\begin{thebibliography}{100}
	\addcontentsline{toc}{section}{Bibliography}
	


\bibitem{1}{S. Perlmutter et al., Astrophys. J. 517, 565 (1999)[astro-ph/9812133].}
\bibitem{2}{A. G. Riess et al., Astron. J. 116, 1009 (1998)[astro-ph/9805201].}

\bibitem{3}{U. Seljak, et al, Physical Review D, 71 (10), 103515(2005).}
\bibitem{4}{ M. Tegmark et al., Phys. Rev. D 69, 103501(2004)[astro-ph/0310723].}
\bibitem{5}{T. Padmanabhan, Phys. Rept. 380, 235 (2003)[hep-th/0212290]. }


\bibitem{6}{S. Weinberg, Rev. Mod. Phys. 61, 1 (1989).}

\bibitem{7}{D. N. Spergelet et al., Astrophys. J. suppl. 148, 175 (2003) [astro-ph/0302209]. }
\bibitem{8}{S. M. Carroll, Living Rev. Rel. 4, 1 (2001)  [astro-ph/0004075].}
\bibitem{9}{P. J. E. Peebles and B. Ratra. Rev. Mod. Phys. 75, 559 (2003)[astro-ph/0207347].}

\bibitem{10}{C. L. Bennett et al., Astrophys. J. suppl. 148,1 (2003)[astro-ph/0302207].}
\bibitem{11}{T. Chiba,  et al., Phys. Rev. D, 62, 023511 (2000). }

\bibitem{12}{Armendariz-Picon, C. et al.,  Phys. Rev. Lett., 85,4438 (2000) ; Phys. Rev. D, 63, 103510 (2001). }

\bibitem{13}{C. Wetterich , Nucl. Phys. B, 302, 668 (1988); yousaf, Z., IlYas, M., Bhatti, M.Z. , Eur. phy. J. plus, 132, 268 (2017).}

\bibitem{14}{B.Ratra, J. Peebles, Phys.Rev.D, 37, 321 (1988).}

\bibitem{15}{A. J. Sen, High Energy Phys., 04, 048 (2002) ; Sharif, M., Eur. Phys. J. Plus, 133(6), 226 (2018).}

\bibitem{16}{T. Padmanabhan, T. R. Chaudhury, Phys. Rev. D, 66, 081301 (2002).}



\bibitem{17}{S. Nojiri, S. D. Odinstov, phys. Lett. B. 562,147 (2003a) ; Phys. Letts. B. 565,1 (2003).}
\bibitem{18}{R. R. Caldwell, Phys. Lett. B. 545, 23 (2002).}

\bibitem{19}{E. Elizalde, S. Nojiri, S. D. Odinstov , Phys. Rev. D, 70, 043539 (2004).}

\bibitem{20}{A. Anisimov, et al., J. cosmol. Astropart. Phys., 06, 006 (2005). }

\bibitem{21} {	A. Kamenshchik, U. Moschella and V. Pasquier, Phys. Lett. B 511, 265 (2001).}

\bibitem{22} {	T.D. Saini, S. Raychaudhury, V. Sahni and A.A. Starobinsky, Phys. Rev. Lett. 85, 1162 (2000).} 

\bibitem{23} {	S. Weinberg, Rev. Mod. Phys. 61,1. (1989) }{G. C. Samanta, Int. J. Theor. Phys., 52, 2303 (2013b).}

\bibitem{24} {	V Sahni, A Starobinsky International Journal of Modern Physics D 9 (04), 373-443 (2000).}

\bibitem{25}{Th P Sotiriou and Va Faraoni, Rev. Mod. Phys. 82, 451 (2010). }

\bibitem{26}{A. De Felice and S. Tsujikawa, Living Rev. Rel. 13, 3 (2010). D. D. Pawar, V. J. Dagwal, Int.J.Theor.Phys.DOI:10.1007/s10773-014.}


\bibitem{27}{P. Horava, D. Minic, Phys. Rev. Lett. 85, 1610 (2000): S. Thomas, Phys. Rev. Lett. 89, 081301 (2002).}

\bibitem{28}{M. Li, Phys. Lett. B 603, 1 (2004) .}
\bibitem{29}{A.G. Cohen, D.B. Kaplan and A.E. Nelson, Phys. Rev. Lett. 82, 4971(1999). }


\bibitem{30}{B. Wang,, Y. Gong, and E. Abdalla, Physics Letters B,  624 (3-4), 141(2005)}

\bibitem{31}{S. Del Campo, J. C. Fabris, R. Herrera, W. Zimdahl,  Physical Review D, 83(12), 123006 (2011).}

\bibitem{32}{ C. Gao, et al., Phy. Rev. D. 79, 043511 (2009).}

\bibitem{33} {J. Zhang, X. Zhang, H. Liu,  Physics Letters B, 659(1-2), 26-33(2008). }

\bibitem{34}{X. Zhang and F. Q. Wu, Phys. Rev. D 72, 043524(2005).}

\bibitem{35}{K. Enqvist, S. Hannestad and M. S. Sloth, JCAP, 0502, 004 (2005).}
\bibitem{36}{D. Pavon and W. Zimdahl, Phys. Lett. B, 628, 206 (2005). }

\bibitem{37}{Z. Chang, F. Q. Wu and X. Zhang , Phys. Lett. B 633, 14 (2006).}


\bibitem{38}{E. N. Saridakis, K. Bamba,  R. Myrzakulov, F. K. Anagnostopoulos, Journal of Cosmology and Astroparticle Physics, 12, 012(2018).}
\bibitem{39}{ E.N. Saridakis, Physical Review D, 102(12), 123525 (2020).}

\bibitem{40}{Wang, Bin, et al., Physics Letters B , 662.1,1-6 (2008)}

\bibitem{41}{Pasqua, Antonio, and Surajit Chattopadhyay, Canadian Journal of Physics, 91.4 ,351-354(2013)}

\bibitem{42}{V. J. Dagwal and D. D. Pawar, Mod. Phys. Lett. A, 33(36), 1850213 (2018).}

\bibitem{43}{A. Pradhan, H. Amirhashchi and B. Saha, Int. J. Theor. Phys., 50(9), 2923-2938(2011).}

\bibitem{44}{A. Pradhan,and H. Amirhashchi, Astrophys. and Space Sci.,332(2), 441-448 (2011). }

\bibitem{45}{D. D. Pawar, G. G. Bhuttampalle, and P. K. Agrawal,  New Astron. 65, 1-6 (2018).}

\bibitem{46}{P. K. Sahoo and B. Mishra, Canadian J. Phys.,92(9) (2014).}

\bibitem{47}{A. P. Billyard and A. A. Coley, Phys. Rev. D 61, 083503 (2000).}

\bibitem{48}{C. Wetterich, Nucl. Phys. B302, 668 (1988). }

\bibitem{49}{C. Wetterich, Astron. Astrophys. 301,321 (1995).}

\bibitem{50}{R. Horvat, Phy. Rev. D. 70  087301 (2004). }

\bibitem{51}{Y. Wang and P. Mukherjee, ApJ, 650, 1 (2006).}

\bibitem{52}{Z. K. Guo. N. Ohta and Y. Z. Zhang, Mod. Phys. Lett. A, 22 ,883 (2007).}

\bibitem{53}{Di Valentino, Eleonora, et al. Classical and Quantum Gravity 38.15,153001 (2021) .}

\bibitem{54}{Li, Miao, et al., Communications in theoretical physics 56.3,525 (2011).}

\bibitem{55}{V. Sahni, T. D. Saini, A. A. Starobinsky and U. Alam, JETP Lett. 77 201 (2003).}

\bibitem{56}{G. R. Farrar and P. J. E. Peebles, Astrophys. J. 604,1 (2004).}

\bibitem{57} {A. Jawad and S. Chattopadhyay,  Astrophys Space Sci. 357, 37 (2015).} 


\end{thebibliography}
\end{document}